\documentclass[a4paper]{jpconf}
\usepackage{graphicx}
\usepackage{amsmath}
\usepackage{amssymb}
\usepackage{amsfonts}
\usepackage{bm}
\usepackage{color}
\usepackage{braket}

\begin{document}
\title{Revisiting Anderson-Higgs mechanism:
application of Lieb-Schultz-Mattis theorem}

\author{Yasuhiro Tada}

\address{Quantum Matter Program, Graduate School of Advanced Science and Engineering, Hiroshima University,
Higashihiroshima, Hiroshima 739-8530, Japan}

\ead{ytada@hiroshima-u.ac.jp}

\begin{abstract}
We consider an electron model of superconductivity on a three-dimensional lattice 
where there are on-site attractive Hubbard interaction and long-range repulsive Coulomb interaction.
It is claimed that fully gapped $s$-wave superconductivity within this model, if present, exhibits 
spontaneous translation symmetry breaking possibly related to a charge order.
Our discussions are based on an application 
of the Lieb-Schultz-Mattis theorem under some physical assumptions.
The inconsistency between the proposed supersolid and experiments 
can impose some constraints on a reasonable choice of a theoretical model.
\end{abstract}

\section{Introduction}
Superconductivity and superfluidity are well understood based on the spontaneous broken U(1) symmetry
of the particle-number conservation.
One of the essential differences between a charged superconductor and neutral superfluid
is that 
there is no Nambu-Goldstone (NG) boson associated with the U(1) symmetry in the former,
while the linearly dispersive sound mode exists as an NG boson in the latter.
The absence of an NG boson in a charged superconductor is often explained by the screening effect of
the long-range Coulomb interaction,
which could be regarded as a variant of Anderson-Higgs mechanism
~\cite{Anderson1958_1,Anderson1958_2,Rickayzen1959,Nambu1960,Wagner1966,Schrieffer}. 
On the other hand, in presence of dynamical (compact) electromagnetic fields,
superconductivity is regarded as an intrinsic ${\mathbb Z}_2$ topological order and
the gapped spectrum is a defining property of this state
~\cite{OS1978,FradkinShenker1979,Hansson2004,Oshikawa2007,Wen}.
It has been naively considered that
these two approaches are consistent and the Coulomb interaction could stabilize the 
topological order.

However, the physics of these two explanations seem to be slightly different;
the gapped NG boson is a consequence of an electric field
coupled to a charge density in the former approach, 
while it is driven by a fluctuating magnetic field coupled to a charge current in the latter approach.
This may be a physically relevant difference in a non-relativistic condensed matter system where 
the electric field and magnetic field can be distinguished~\cite{com1}.
Furthermore, in the former, 
the Meissner effect
can be described with use of the Maxwell equation for a magnetic field
even without the long-range Coulomb interaction, which would suggest that the Meissner effect and
the gapped NG boson are independent phenomena.
In contrast, these two phenomena take place simultaneously in a gauged model,
which is common for general gauge theories.
Besides, there could be experimentally measurable differences 
between these two approaches in some superconductors
such as a superconductor whose normal state is a semimetal or an insulator with a vanishing Fermi surface
and a purely two-dimensional superconductor at zero temperature. 
In such a superconductor, 
screening effects may be too weak to gap out the NG boson in the charge screening approach,
while a gapped spectrum will be by definition robust in a gauged model.

In this study, we revisit this subtle problem based on the Lieb-Schultz-Mattis theorem for low energy
spectra and ground state degeneracy (GSD) for general quantum many-body systems~\cite{LSM1961,Tada2021}.
We argue
that a conventional $s$-wave superconductor described by an attractive Hubbard model
with a long-range Coulomb interaction has GSD $\geq q$ when the particle filling per unit cell
is fractional $\rho=p/q$ with coprime integers $p, q$.
The ground states exhibit translation symmetry breaking possibly corresponding to a charge-density-wave
formation with period $q$,
and consequently, the superconducting state will be a supersolid.
The inconsistency between the proposed supersolid and experiments 
can impose some constraints on a reasonable choice of a theoretical model.

.

\section{Possible supersolid in electron model of superconductivity}
The aim of this study is to investigate possible outcomes of a reasonable model of superconductivity
based on the well established experimental fact that an $s$-wave superconductor is fully gapped.
This in turn can impose some constraints on a model choice.
For this purpose,
we consider the following Hamiltonian defined on a cubic lattice of a linear size $L$
with the periodic boundary condition,
\begin{align}
H=-\sum_{\langle ij\rangle\sigma} tc^{\dagger}_{i\sigma}c_{j\sigma}-U\sum_{i}n_{i\uparrow}n_{i\downarrow}
+\frac{1}{2}\sum_{i\neq j}V_{ij}(n_i-\rho)(n_j-\rho),
\label{eq:H1}
\end{align}
where $\langle ij\rangle$ is a pair of nearest neighbor sites and $\sigma=\uparrow,\downarrow$
is the spin index.
$c_{i\sigma}$ is an annihilation operator of an electron at site $i$ and spin $\sigma$,
and $n_{i\sigma}=c^{\dagger}_{i\sigma}c_{i\sigma}, n_i=\sum_{\sigma}n_{i\sigma}$. 
The second term is an attractive interaction arising from e.g. phonons,
and the third term is the repulsive Coulomb interaction on a lattice,
$V_{ij}=V/|i-j|$ for $i\neq j$.
(More precisely, the attractive term should be regarded as a sum of a phonon mediated interaction
and on-site repulsive Coulomb interaction.)
The charge density is measured from the average density $\rho=p/q$ per unit cell with coprime integers $p,q$.
The total number of electrons is $N=\rho L^d$ with $d=3$ in the present model.
Equation~\eqref{eq:H1} is considered to be one of the simplest models for charged superconductivity and 
one can generalize the discussions to more complex systems with local orbitals on other lattices.
Note that although our argument is applicable also to a two-dimensional system,
we focus on three-dimensional cases which are more non-trivial since a plasmon is not gapped 
for a normal state in pure two dimensions and screening effects are expected to be similarly weak
in a purely two-dimensional superconductor at zero temperature~\cite{Ando1982}.
We simply assume that eigenvalues of the Hamiltonian are extensive and the system has a well-defined
thermodynamic limit. 
Throughout our discussion, we focus on zero temperature, 
and strictly keep the U(1) particle-number conservation since we
consider an isolated superconductor.
In this framework, the U(1) symmetry is not broken although there can be a corresponding long-range order
due to the attractive interaction $U$,
and the ground state is a Schr{\"o}dinger's cat state in terms of the global U(1) phase 
of the Cooper pair order parameter~\cite{KT1994,Tasaki2019,Tasaki}.
Any gauge invariant quantity can be correctly evaluated within
this fixed particle-number description, and differences are seen only for gauge dependent quantities
such as the order parameter which is not an observable.

In absence of the Coulomb interaction ($U\neq0,V=0$), Eq.~\eqref{eq:H1} is reduced to the well-known attractive 
Hubbard model,
and the corresponding ground state is expected to be a spatially uniform $s$-wave 
superfluid~\cite{Lieb1989,Auerbach,com_eta,Shen1993,Tian1994,Shen1998}.
On the other hand, when $U=0,V\neq0$, the system would be either metallic or charge ordered 
as seen in the previous studies of extended Hubbard models~\cite{Auerbach,Giamarchi,aHubbard1,aHubbard2}.
Naively, one would expect that superconductivity and a charge order can coexist to form a supersolid state
at some parameter region with both non-zero $U,V$.
For example, in the strong coupling limit ($t\ll U$ and $V\neq0$),
ground states for $\rho=2/3$ would be
\begin{align}
\ket{\tilde{\Psi}_0}&=
\ket{\cdots,\uparrow\downarrow,0,0,\uparrow\downarrow,0,0,\cdots}
+\rm{(corrections)},\nonumber\\
\ket{\tilde{\Psi}_1}&=
\ket{\cdots,0,\uparrow\downarrow,0,0,\uparrow\downarrow,0,\cdots}
+\rm{(corrections)},
\label{eq:phi_guess} \\
\ket{\tilde{\Psi}_2}&=
\ket{\cdots,0,0,\uparrow\downarrow,0,0,\uparrow\downarrow,\cdots}
+\rm{(corrections)},\nonumber
\end{align}
where they are related by the translation operator as ${\mathcal T}_x\ket{\tilde{\Psi}_0}=\ket{\tilde{\Psi}_1},
{\mathcal T}_x\ket{\tilde{\Psi}_1}=\ket{\tilde{\Psi}_2}$, and ${\mathcal T}_x\ket{\tilde{\Psi}_2}=\ket{\tilde{\Psi}_0}$.
In this case, 
the on-site $\uparrow\downarrow$ Cooper pair is regarded as a  boson which
has an effective hopping $\sim t^2/U$ in the second order perturbation when $U\gg t$,
and a supersolid of the effective bosons could be stabilized by the interaction $V$
~\cite{bSS1,bSS2,bSS3}.
This is a strong coupling picture for $U\gg t$ corresponding to a BEC regime.
Although basic structures of these candidate ground states are reasonable, 
it is highly non-trivial whether or not the true ground states can indeed exhibit 
both superconductivity and a charge order simultaneously
in some parameter region especially for small $U$ corresponding to a BCS regime.
For example, it would be possible that
a superfluidity driven by $U$ and a charge order stabilized by $V$
are separated by a phase transition and there is no coexisting region.

Here, we do not directly
examine existence or absence of a superconducting state in our model with both non-zero $U,V$,
but instead we introduce some physical assumptions which are necessary to describe desired properties of
superconductivity.
We first focus on low energy states in the attractive Hubbard model at $V=0$
and later take the Coulomb interaction into account.
Now we simply assume that there is a long-range order of the spin-singlet Cooper pairing in the ground 
state of the attractive Hubbard model,
$\bra{\Psi_0}{\mathcal O}^{\dagger}{\mathcal O}\ket{\Psi_0}\sim (L^d)^2$.
The bulk order parameter
is defined as ${\mathcal O}=\sum_ic_{i\uparrow}c_{i\downarrow}$.
In this case, single-particle fermion excitations and spin excitations will be gapped and thus
it is natural to assume that there are only two kinds of low energy states in 
the $N$-particle Hilbert space ${\mathcal H}_N$,
namely, the ground state $\ket{\Psi_0}$ and states with NG excitations.
It is known that 
the NG boson is related with collective particle density excitations, where the variational states (up to normalization)
\begin{align}
\ket{\Phi_k}=N_k\ket{\Psi_0}, \qquad N_k=\sum_j n_je^{ikr_j}
\end{align}
are good approximations of the exact excited states
~\cite{Wagner1966}. 
According to the NG theorem, the variational energy of these states are 
$\simeq {\rm Const}\cdot k\sim L^{-1}$
measured from the ground state energy $E_0$.
It is important to remark that there are other low energy states associated with the U(1) symmetry breaking 
if one extends the Hilbert space to
include different particle-number sectors~\cite{KT1994,Tasaki2019,Tasaki}.
In this case, 
the entire Hilbert space is ${\mathcal H}=\bigoplus_{N=0}^{2L^d} {\mathcal H}_N$ and 
the tower of states has energies $\sim L^{-d}$ measured from $E_0$.
The tower of states is well approximated by the variational states 
${\mathcal O}^m\ket{\Psi_0}$ and $({\mathcal O}^{\dagger})^m\ket{\Psi_0}$ up to normalization.
However, these states belong to ${\mathcal H}_{N-2m}$ and ${\mathcal H}_{N+2m}$ respectively,
but not to ${\mathcal H}_{N}$ which contains $\ket{\Psi_0}$.
As mentioned before, these states are not relevant in the present study for an isolated 
superconductor with the particle-number conservation law.

Next, we consider the full Hamiltonian with both non-zero $U, V$ under the assumption
that the system is superconducting~\cite{com3}.
To have a constraint on possible low energy spectrum and GSD, 
we use the Lieb-Schultz-Mattis theorem for a long-range interacting system recently proved by the author
~\cite{LSM1961,Tada2021}. 
For the proof of the theorem, 
variational low energy states $\ket{\Psi_1},\cdots,\ket{\Psi_{q-1}}$
are constructed with use of translation symmetry and U(1) symmetry,
and these states including $\ket{\Psi_0}$ are distinct each other when
the filling per unit cell $\rho=p/q$ is fractional.
The existence of such states implies that
only the following two possibilities are allowed in the present model.
\begin{enumerate}
\item There are gapless excitations above the ground state(s). 
\item There is a gap between the ground state sector
and excited states, and the ground state sector has GSD $\geq q$.
\end{enumerate}
This is a non-perturbative statement which is valid for any parameters,
but we cannot fully identify which one will be the case in the present system.
However, as was discussed in the introduction,
the long-range Coulomb interaction can lead to the screening effect and the NG boson will be gapped out.
Therefore, there are no low energy excited states in ${\mathcal H}_N$ above the ground states,
and the first possibility (existence of gapless excitations) is excluded.
The remaining possibility is that there are at least $q$-hold degenerate ground states with a gap above them.
We stress that the system cannot be a simple uniform ${\mathbb Z}_2$ topologically ordered state,
because GSD $\geq q$ is filling dependent and is clearly distinguished from the ${\mathbb Z}_2$ topological 
degeneracy which depends only on the global topology of the underlying manifold on which the Hamiltonian 
is defined.
Besides,
these degenerate ground states are different from the tower of states since the former belong to ${\mathcal H}_N$.
The remaining resolution is that they are related to translation symmetry breaking~\cite{Koma2000,Oshikawa2000}.
Indeed, each ground state is essentially an eigenstate of the translation operator, 
${\mathcal T}_x\ket{\Psi_n}=e^{iP_n}\ket{\Psi_n}$ with $P_n=2\pi n/q$,
and one can construct physical ground states (up to normalization) from these cat states
~\cite{Oshikawa2000},
\begin{align}
\ket{\tilde{\Psi}_n}=\sum_{n'=0}^{q-1}e^{iP_nn'}\ket{\Psi_{n'}}.
\end{align}
Then, the translation symmetry is spontaneously broken in 
the physical ground states and they are related each other as
${\mathcal T}_x\ket{\tilde{\Psi}_n}=\ket{\tilde{\Psi}_{n+1}}$ (mod $q$) just like the naive guess Eq.~\eqref{eq:phi_guess}.
From the previous studies for extended Hubbard models~\cite{Giamarchi,aHubbard1,aHubbard2},
it is reasonable to suppose that
the translation symmetry breaking in the present Hamiltonian is related to a
charge-density-wave order,
although other non-uniform states such as a Fulde-Ferrell-Larkin-Ovchinnikov state are not completely excluded.
To summarize, the resulting states are superconducting and break translation
symmetry simultaneously possibly corresponding to a charge order. 
This means that if the Hamiltonian Eq.~\eqref{eq:H1} can describe the desired properties of superconductivity
as we have assumed, the outcome is that the ground state is generally a supersolid.
We stress that our argument is based on the non-perturbative Lieb-Schultz-Mattis theorem, and thus
the statement holds for any parameters as long as the assumptions are met.
Note that 
if the two approaches for the gapped NG boson based on the screening effect and gauge fluctuations respectively
are consistent as mentioned in the introduction,
the supersolid state will also be simultaneously a ${\mathbb Z}_2$ topologically ordered state.

\section{Discussion and summary}

In the previous section, we have shown that  a system described by the electron Hamiltonian
with both a local attractive interaction and long-range Coulomb interaction
will exhibit a supersolid state.
Our argument is non-perturbative and general,
although it is just qualitative and magnitude of a charge order could be tiny 
in a realistic parameter region.
Although we have introduced several assumptions (U(1) long-range order,
no gapless excitations other than the NG boson at $V=0$, and gapping NG boson by the Coulomb interaction)
so that the argument goes step by step,
they are combined into simpler assumptions that the Hamiltonian Eq.~\eqref{eq:H1} can
describe superconductivity and has a gapped spectrum~\cite{com2}.
Of course, from a purely theoretical point of view, 
these assumptions are non-trivial and should be examined,
but a breakdown of the assumptions implies that the chosen model is not appropriate for superconductivity.
Our proposal of a supersolid state sharply contradict many existing experiments, although there are some
superconductors with charge orders.
In this view point, our result implies that the simple model of superconductivity Eq.~\eqref{eq:H1} may not 
be a fully correct choice.
One can add some other density-density type interactions for which the Lieb-Schultz-Mattis theorem
is still applicable~\cite{Tada2021},
and also our statement would be perturbatively robust to more general short-range interactions 
since the system is assumed 
to be gapped.
In this way,
the inconsistency between the proposed supersolid and experiments
can impose some constraints on a reasonable choice of a theoretical model.

As mentioned in the introduction, superconductivity can be alternatively described by a gauged model
such as the compact U(1) Higgs model corresponding to the Ginzburg-Landau theory.
In this phenomenological model, a spatially uniform ${\mathbb Z}_2$ topological ordered state is realized.
Similarly, superconductivity in a gauged electron model 
coupled to a dynamical electromagnetic field
is not characterized by a local order parameter~\cite{TK2016,TK2018}.  
However, to our best knowledge, 
there is no theoretical study on a possible charge order in such a gauged electron system. 
One may naively expect that a uniform state is stable since the interactions are short ranged in the 
${\mathbb Z}_2$ ordered phase~\cite{Wen}, but a charge ordered state could be stable as well since it is gapped. 
Besides, there is another subtle issue for gauged models about compactness of the U(1) 
electromagnetic field.
It is known that a non-compact U(1) Higgs model with a gauge fixing
can show a long-range order of a bosonic condensation
and the correlation function decays algebraically implying existence of an NG boson~\cite{KennedyKingPRL}.
Although one would naively expect that this candidate NG boson is decoupled from the rest of the system and thus
cannot be observed through local gauge invariant quantities,
such excitations could contribute to thermodynamic quantities such as specific heat,
which is inconsistent with experiments.
Nevertheless, we believe that gauged models are promising models, since they necessarily include magnetic fields
and coupling to charge currents.
Especially in compact gauge theories, gapping photons (Meissner effect) and gapping NG bosons occur at the same time  
once a coupling between gauge and matter fields is introduced,
in contrast to a pure electron model with Coulomb interaction where these two phenomena take place separately.
This is consistent with the conventional Ginzburg-Landau description of superconductivity which
well explains experiments~\cite{comGL}.
Therefore, we expect that superconductivity can be better described by a gauged model.

The problem discussed in the present study is a highly subtle issue, and indeed the two theoretical approaches 
for the absence of the NG boson have been thought basically equivalent.
The key is a coupling to the U(1) gauge field, which is to some extent common for both approaches.
However, as discussed above, 
there is no gauge degrees of freedom and no magnetic field
in a Coulomb interacting model which takes only an electric field into account (in this sense the model is non-gauged),
while dynamical electromagnetic fields are fully coupled to electrons in a gauged model.
This may bring a clearer difference in their outcomes than the charge order proposed in the present study.
Our crude discussion is just a single step, 
and further investigation will be necessarily for this subtle but fundamental problem.

\ack
We are grateful to Y. Yao, M. Oshikawa, and A. Koga for valuable discussions.
This work was supported by JSPS KAKENHI Grant No. JP17K14333.

\section*{References}
\bibliography{ref_iop.bib}

\begin{thebibliography}{10}
\bibitem{com1}
For example in a three-dimensional metal, the electric field is short ranged
  due to a screening effect, while the magnetic field is long ranged.
\bibitem{com3}
We do not explicitly assume the U(1) long-range order at $V\neq0$, because the
  system might be a gapped ${\mathbb Z}_2$ topologically ordered state where
  any correlation functions are short ranged. Instead, we say that an electron
  system is superconducting when it shows Meissner effect.
\bibitem{com2}
Our argument can be greatly simplified if we simply assume that the spectrum of
  the Hamiltonian Eq.~\eqref{eq:H1} is gapped so that it is consistent with the
  experimental results of the gapped superconducting nature. By applying the
  Lieb-Schultz-Mattis theorem to the Hamiltonian, we immediately conclude that
  the ground states are at least $q$-hold degenerate under this assumption.
  This implies a supersolid state when the system is superconducting, as
  discussed in the main text. Although this is just a straightfoward
  application of the Lieb-Schultz-Mattis theorem, it has not been discussed so
  far to the best of our knowledge.
\bibitem{comGL}
Note that a static Ginzburg-Landau theory with no time dependence should
  provide a sufficient description of the Meissner effect and gapped NG boson,
  since they are equilibrium properties of superconductivity. In this
  description, no electric field is necessary.
\end{thebibliography}

\end{document}